# ROAD QUALITY ANALYSIS BASED ON COGNITIVE INTERNET OF VEHICLES (CIOV)

Hamed Rahimi[1] and Dhayananth Dharmalingam[2]

Université de Lyon, France

APRIL 2020

---

[1] Hamed.Rahimi@etu.univ-st-etienne.fr

[2] Dhayananth.Dharmalingam@etu.univ-st-etienne.fr

# CONTENTS

# ABSTRACT

This research proposal aims to use cognitive methods to analyze the quality of roads based on the new proposed technology called Cognitive Internet of Vehicles (CIoV). By using Big Data corresponding to the collected data of autonomous vehicles, we can apply cognitive analytics to a huge amount of transportation data. This process can help us to create valuable information such as road quality from an immense volume of meaningless data. In this proposal, we are going to focus on the quality of roads for various business and commercial purposes. The proposed system can be used as an additional service of autonomous car companies or as a mobile application for ordinary usages. As a result, this system can reduce the usage of resources such as energy consumption of autonomous vehicles. Moreover, this technology benefits the next-generation of self-driving applications to improve their QoS.

# I. INTRODUCTION

The combination of AI and the Internet of things, in the context of autonomous vehicles, has brought a new concept called the Internet of Vehicles (IoV). IoV uses new computing methods such as cognitive computing to improve the QoS of the next-generation of self-driving car applications. From the social point of view, these systems can reduce the usage of energy and resources, which are limited and valuable for us. However, these systems also have brought us new technical issues such as managing, collecting, and transmission of data.

## 1. Motivation

Due to the accelerated growth of autonomous vehicle applications, we can simply imagine that soon we would deal with huge, varied and numerous transportation data. Therefore, it is obvious we can consider that situation as an opportunity to utilize these data to generate meaningful and valuable information for various purposes. In this regard, we explore the quality of roads that can be used as an additional service of autonomous vehicles. These systems can reduce the energy consumption of autonomous vehicles and improve the QoS of the next generation of self-driving applications.

## 2. Objective

The Fourth Industrial Revolution (4IR) or Industry 4.0 [1] has been represented in automation to refer to various important technology innovations such as Cyber-Physical Systems (CPS) [2] and Artificial Intelligence (AI) [3]. The Internet of Things (IoT) [4] as of the important parts of 4IR has been becoming extensively demanding. According to Statista research [5], it is predicted that more than 90 percent of cars by 2025 will be connected to the internet. On the other hand, AI has been increasing the intelligence of things in contrast to the natural intelligence displayed by humans. AI is a demonstrated machine Intelligence that includes learning, reasoning and self-correction. Autonomous Vehicle (AV) [6], or self-driving vehicle, is a concept referring to the vehicles that through the ability to sense surroundings can operate by itself and do its necessary tasks without any human interruption. The combination of AI and the Internet of things, in the context of autonomous vehicles, has brought a new concept called the Internet of Vehicles (IoV) [7]. The Internet of Vehicle (IoV) is a distributed network that consists of connected vehicles and uses new computing methods such as cognitive computing to improve the QoS of the next-generation of self-driving car applications. The architecture of CIoV has been shown in Fig-1. In this regard, there is a new concept called Cognitive Internet of Vehicle (CIoV) [8] that has been proposed as a high-level solution to improve the cognitive intelligence of IoV. However, CIoV faces various challenges that we will explain in the Section III. By using the Cognitive method, CIoV can estimate the quality of roads and show an efficient path from a current location toward a destination.

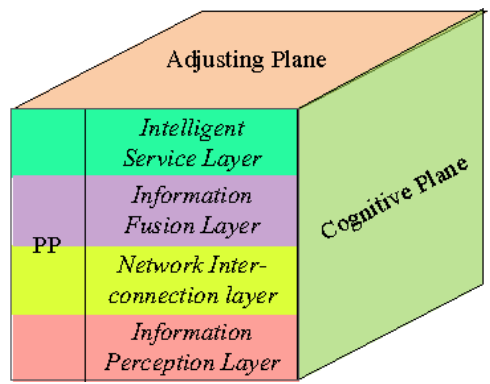

Fig-1 The architecture of CIoV [17]

### 3. State of Art

Roads are the general way of transportation. Therefore, the sustainability of road quality is required for the safety of the people. Nowadays, common navigation systems used by people (such as Google Maps) assists us to find a faster route to our destination. Regularly, this process is based on the speed of vehicles in a particular area. However, road quality does not define only by speed. Prapulla et al. [9] proposed a system that analyzes the road quality of an area based on applying an accelerometer sensor and GPS to make a map of roads. Badurowicz et al. [10] proposed a cloud computing system designed for observing the status of roads by processing the car's acceleration and position data acquired by smart mobile devices implemented on the IBM BlueMix platform. in [11], Astartita et al. proposed the mobile anomaly-sensing systems where data from mobile agents processed on a central server after the data acquisition process. In [12], Khelil et al. explored D2D and researched its suitability to the safety-critical Internet of Vehicles (IoV) applications, such as cooperative or self-driving. In [13], Chowdhary et al. implemented two famous algorithms Ant Colony Optimization and Particle Swarm Optimization to enable coordinated dynamic route customization among the vehicles for overall optimal traffic management.

## II. RESEARCH DESIGN AND METHODS

### 1. Hypothesis

Current maps and route suggestion applications are still missing some important features and information which is important to people. Enabling road quality information on the map would be helpful to the users. "Road quality analysis system" would be developed by considering the following hypothesizes. It is hypothesized that future technology will enable different opportunities for developers with innovation and invention, which can help to implement several features to the map. Existing maps can be enabled with road quality information, warning, and alternative route suggestion by analyzing different factors. This advance route suggestion system will be calculated based on characterized and structured information. Current tech companies provided a promising opportunity to build powerful smart inventions. As it is mentioned in Tesla (Tesla Corp, 2019), the product of tesla will be enabled with edge and curve detection analysis for the great experience of the users. The data can contain many features and hidden information. According to the (Pardenilla, 2018), a self-driving vehicle will identify and inform to the owner regarding the problems and condition of each part of a vehicle based on the road it has driven through. That shows that autonomous vehicles use different technologies to understand the road condition up to a certain level. Accordingly, it is expected that future technologies will retrieve and store various information about the environment and road surface that will further be processed by the proposed system to enable additional features on the map. These additional characterized features will helps to detect better route decisions to the users. Hence, the users can be benefited from this feature when selecting an appropriate route for their drive. In addition, this route mapping decision would help to select the safest, fastest and smoothest drive. The model of proposed research has been shown on Fig 2.

## 2. Requirement

First, we need to collect the various types of data from various systems of vehicles to process and using them to train the proposed system. Therefore, vehicles need to be connected through a network. Internet of Vehicles (IoV) based on IoT connects the vehicles through the internet. On the other hand, the proposed system must have access to the collected data of vehicles in a certain district. Therefore, data from different systems have to interoperate with the data of other systems. In other words, we need ways to semantically analyze and systematically extract information from datasets, which are too large and complex. In this regard, Semantic Intelligence and Big Data can help us to access, collect and analyze the data. Second, we need to process the pre-processed data to train our system and update it frequently with new entries. Therefore, we use cognitive algorithms to process data and train our system. The cognitive algorithm produces a concept as its output, it means that cognitive learning explains thinking and differing mental processes, which have better comprehensive performance. Moreover, it explains how they are influenced by internal and external factors to produce learning in individuals.

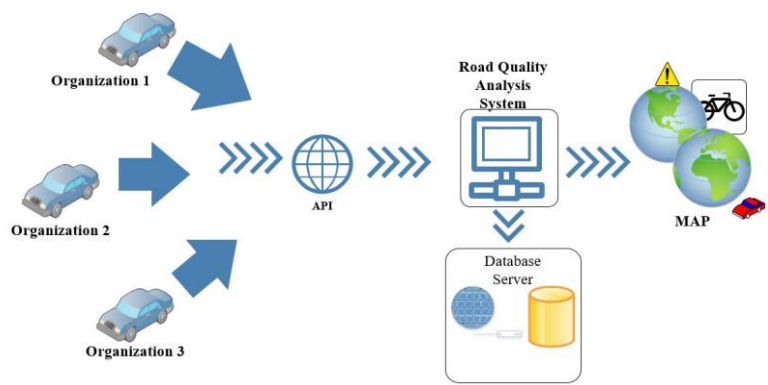

Fig 1- The proposed model

# III.   CHALLENGES

## 1.   Technical Challenges

There are various technical issues in implementing the research in practical areas.

### 1.1 Interoperability

This project assumed to utilize data from different future technologies such as the tesla autonomous project (Tesla Corp, 2019), google smart city (Edwards, 2019) and so on. That will result in different formatted and structured data from different sources. Standardizing and unifying data for one particular format would require additional work force and increase the affordability of the project. It requires some additional effort to utilize different data sources in order to produce the most accurate and more reliable result.

### 1.2 Content negotiation and Semantic Interoperability

Required data have been expected to acquire from other organizations. In this case, content negotiation would be a big deal. Organizations will never accept maintaining a standard suggested by some other people. They will utilize technologies and content formats based on their requirement, which is obviously accepted. Therefore, the development team should process many data sources and formats. Therefore, the additional workload is necessary for the development team to parse, read and process data from the different content formats.

### 1.3 The ambiguity of data format

Since the data has been planned to collect from different organizations, there will be a definite problem with the "ambiguity of data". Different organizations will follow different standards and techniques to name key values. Even though some IT leaders try to implement unified standards (Google, Microsoft, Yahoo and Yandex, Schema.org, 2019) to reduce data ambiguity, still it is in question and not widely used.

### 1.4 Accuracy of the result

This quality of the road will be analyzed by collecting data from different organizations and technologies. The data can be collected for different purposes by other organizations. Hence, road quality information will be a hidden feature of this dataset. Selecting the proper data mining technique and approach will be a great challenge. In addition, achieving 100% accuracy in the result will be a big challenge for the development team.

### 1.5 Extracting underlying information

In the beginning, it is necessary to choose suitable sensors for capturing data. Choosing an appropriate sensor is an important task for the development group to extract the required information for the analyses. The Internet of Vehicles will be enabled with many sensors. The development team should capable to extract the required information from captured data. Because all required information cannot be captured directly with high tech sensors. For instance, a proximity sensor will help to capture the condition of the road surface and computer vision will help to analyze the edges of the road. Both data can be processed further to identify speed limit, danger curve indication and so on. Likewise, extracting required information from other purpose data and selecting the most efficient and effective approach would be a challenging task for the development team.

## 2. Social Challenges

There are various social challenges in the deployment of IoV applications.

### 2.1 Acceptance

As it is expected to gather data from different organizations, the main problem would be acceptance. Mostly, organizations will not be convinced or accept to share data with third-party organizations. In addition, in some cases, it might be required to implement some additional sensors or embedded systems to the autonomous vehicle. Organizations might not accept to perform such a task to their innovations. Thus, the acceptance question will be a big challenge for this project.

### 2.2 Legal Obligations

Some countries may not accept to share road quality data due to different negative impact of the country. Therefore, getting legal acceptance from such countries will be a challenging task in order to enable this map feature to the relevant country.

### 2.3 Ethics

The ethical aspect of data and privacy is highly concern in the digital era. In addition, various law and legal formation enforce appropriate usage of information. Proper security features required protecting and transfer data more secure manner. In addition, information utilization should be transparent and assure the legal usage of information. Application and analyzing techniques should be implemented by concerning ethical factors and legal requirements. Otherwise, the application can be banned from some countries by considering the misuse act (UK Gov, 2012)

### 2.4 Environmental protection

Environmental protection is also a considerable social challenge. IoV systems should consider the environmental aspects of autonomous vehicle applications to protect the environment and to reduce the usage of resources such as time and petrol. Addressing this aspect can help many people to save more time and money. Moreover, it helps to prevent various potential traffic hazards, vehicle damages, danger accidents due to poor road condition and so on.

### 2.5 Service providing

IoV systems determine critical driving spots by measuring the speed, the road quality, slope detection of tires, and the width of the vehicle-driving lane in the road. Those data can be processed to implement warning messages and indications to the users, which can alert the driver to increase awareness about driving through certain areas. That helps to decrease many hazards and inconvenience situation happening in high ways.

### 2.6 Regulatory

Different countries legitimate with different traffic rules and traffic systems. Hence, this autonomous project should adhere to those rules and regulations. A small error in the deployment could result in major accidents in a real-world application. Therefore, developing a general system, to draw a map and to estimate the quality of roads, is a challenging task. To address this challenge, proper analysis and research should be done with real world experimental before the live deployment to the public.

## IV. EVALUATION

### 1. Overview

In order to evaluate the performance of proposed research, firstly we apply the proposed algorithm to real-world data to simulate a map with road quality. Based on the drawn map, we try to predict the best way of reaching from an initial point to a destination.

At first, the application will be demonstrated in a limited area with a small coverage range. This testing process will be done by the development team and no user will be involved in this test. Inputs, results, and accuracy of the results will be evaluated in this stage. Necessary fixes and changes will be made before the release of the beta application. Next, a beta application will be released to the public. The beta version of the application will be released with limited features to get user reviews. Changes and additional features will be released one by one in the meantime based on the reviews of the users. It helps to improve and analyze the result more efficiently. Therefore, can able to provide great reliable and most accurate route suggestion apply to the users. The initial release of the application will be done in selected countries. Therefore, the evaluation process will be easier, maintenance and upgrading processes also can be done easily. Finally, we compare our results to other methods to clarify the differences and advantages of the proposed algorithm with the existing features. Effectiveness of the Warning and other road signals will be evaluated with user reviews. In the future, the algorithms and approaches will have collaborated with leading map applications in order to provide the most advanced map features to the users, especially for driving. In future research, we will consider more complex data structures to improve the QoS of the algorithm.

## 2. Impacts and Benefits of the Project

Road quality analysis system will help to add additional features to the existing map by analyzing the quality of the road. This feature is also an important point in a map and it must be considered when selecting a route. The article "The impact of road improvements on road safety and related characteristics" (JohnGichaga, 2017), says that the majority of accidents in developing countries caused by the condition of the road. The proposed application assures the safest drive to the user with a different form of indication. This application will create speed limits, warning signs, danger signs in the appropriate location on the map, by mainly considering the quality of the road. Therefore, the user will be warned and informed in advance, which helps to increase awareness. Therefore, the risk of road hazards will be reduced possibly. Some road conditions can cause dangerous accidents. For example, potholes on the road surface will severely influence and change the direction of a moving car. This can cause a dangerous situation when a vehicle moves on potholes at a certain speed. So enabling a warning system with proper indication will extremely reduce unexpected hazards in the street. According to (FWHA, 2018), 21% of crashes were registered due to weather conditions, 418,005 people were injured between 2007-2016 in the US. Some roads in a certain area were not well designed with appropriate road signs. That causes a difficult situation for the driver. The proposed application will dynamically insert road signs and indications based on collected data from previous driven vehicles on selected roads. In addition, it will analyze and calculate the speed limit by considering many factors such as the condition of the road, turning capacity of the vehicle, the legal requirement of the speed limit on the selected road, weather season, and previous accidents in a certain distance. These signs and advance route mapping systems will help to plan the safest drivers in the dangerous areas. In addition, that helps to select the fastest and safest route with different options that can be selected by users based on their vehicle type. It can reduce accidents and unexpected hazards caused by road quality and poor road signs. This system can be utilized by tire manufactures to make business decisions. For example, based on the road condition, manufacturing companies can decide what kind of tires should be delivered to the selected area. In addition, road development and enhancement activities can be planned and decided by relevant government authorities with the information provided in the proposed application. Hence, the "Road quality analysis" application will create many positive impacts and benefits for the stakeholders. Enabling this feature to the public will helps to save people from road hazards and dangers.